\documentclass[referee]{aa-5-mpa}
\usepackage{graphicx}
\usepackage{amssymb}
\usepackage{txfonts}
\usepackage{natbib}

\usepackage{natbib}
\bibpunct{(}{)}{;}{a}{}{,}  

\newcommand{\tralpha}{3\,\alpha}
\newcommand{\Iso}[2]{^{#1}{\rm #2}}
\newcommand{\nbottle}{\Iso{14}{N}(p,\gamma)\Iso{15}{O}}
\newcommand{\msun}{M_\odot}
\newcommand{\lsun}{L_\odot}
\newcommand{\teff}{T_\mathrm{eff}}

\newcommand{\rbox}[1]{\raisebox{1.2ex}[0pt]{#1}}

\begin{document}

\title{Influence of two updated nuclear reaction rates on the
evolution of low and intermediate mass stars}

\author{A.~Weiss\inst{1} \and A.~Serenelli\inst{2,1} \and
A. Kitsikis\inst{1} \and H.~Schlattl\inst{1} \and
J.~Christensen-Dalsgaard\inst{3}}

\institute{Max-Planck-Institut f\"ur Astrophysik,
           Karl-Schwarzschild-Str.~1, 85748 Garching,
           Federal Republic of Germany
           \and
           Institute for Advanced Study, Einstein Drive, Princeton, 
	   NJ~08540, USA
           \and
           Institut for Fysik og
              Astronomi, Aarhus Universitet, Bygn. 520, Ny Munkegade,
	      DK-8000 Aarhus C, Denmark
           }

\offprints{A.~Weiss;\\ (e-mail: weiss@mpa-garching.mpg.de)}
\mail{A.~Weiss}

\date{Received; accepted}

\authorrunning{Weiss et al.}
\titlerunning{Stellar models with revised reaction rates}

\abstract{Two key reactions of hydrostatic nuclear burning in stars
have recently been revised by new experimental data -- the $\nbottle$
and $\tralpha$ reaction rates. We investigate how much the new rates
influence the evolution of low-mass, metal-poor and metal-free, stars
and of an intermediate-mass star of solar-type
composition. We concentrate on phases of helium ignition or thermally
unstable helium burning. Our global result is that the new $\tralpha$
rate has no significant influence on such stars, but that there is a
noticeable, though small effect of the new $\nbottle$ rate, in
particular on the core helium flash and the blue loop during core
helium burning in the intermediate-mass star.
}
\maketitle
\clearpage

\section{Introduction}

\begin{figure}[bh]
\centerline{\includegraphics[draft=false,scale=0.5,angle=90]{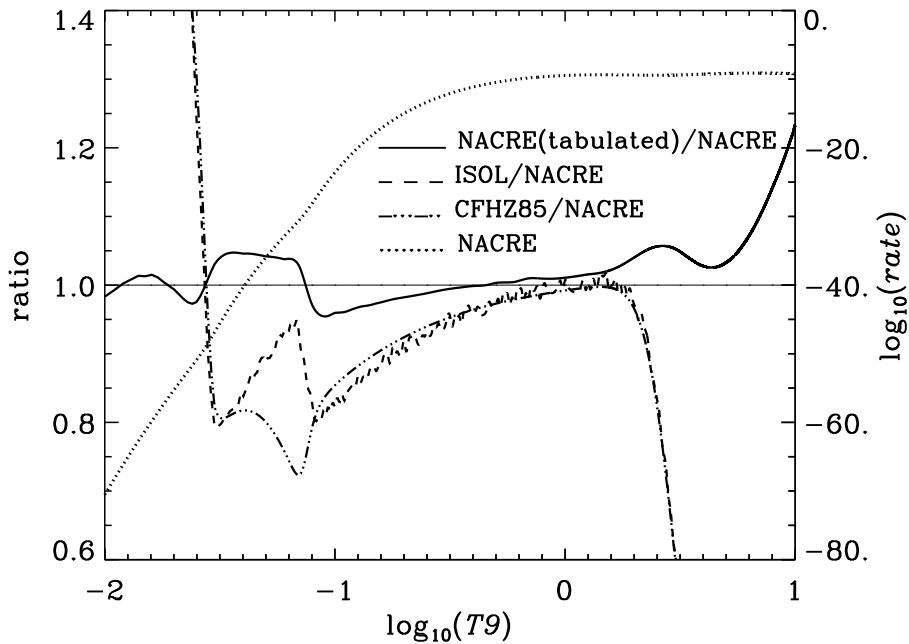}}
\caption[]{$\tralpha$ reaction rates relative to the NACRE analytic one
\citep{nacre:99}. Shown are the tabulated
NACRE, CFHZ85 \citep{CFHZ:85} and ISOL rates
\citep{tralpha:05}. The dotted line is the
absolute value of the reference NACRE rate (scale on right axis).}
\protect\label{f:1}
\end{figure}

The triple-alpha reaction is one of the key nuclear reactions for the
synthesis of the elements in stars and is also the main energy
source during helium burning. The reaction rate is dominated by
resonances, the best known being the one at 7.65~MeV, theoretically
predicted by \citet{hoyle:54}, but there is
considerable interest in determining all resonances with high
precision. Recently, \citet{tralpha:05} reported new measurements
concerning resonances of $\Iso{12}{C}$ with 3 $\alpha$-particles obtained from
$\Iso{12}{C}$-decay experiments. In particular, they found a dominant 
resonance at $\simeq 11$~MeV, while they did not confirm another at
9.1~MeV reported previously \citep{nacre:99}. 
The new reaction rate  (called the ``ISOL'' rate in the following)
deviates from the one published by \citet[NACRE]{nacre:99}: at
temperatures between $2.5\cdot 10^7$ and  $10^8$~K it is
between 7 and 20\% lower, but below $2.5\cdot 10^7$~K up to 50\%
higher. The largest change occurs for $T > 3\cdot 10^9$~K, where the
new rate is increasingly lower by up to one order of magnitude
(Fig.~\ref{f:1}).

In the present paper we consider three
cases of stellar evolution where a modified $\tralpha$ rate might
influence the models: the core helium flash in
low-mass metal-poor stars, the core and shell helium burning in
intermediate-mass stars of solar-like composition, and helium burning
in metal-free low-mass stars. This mass range is interesting,
because of the high temperature sensitivity during core and shell
flashes, where instabilities amplify even small temperature
variations.

Coincidentally, another key reaction has been re-determined recently by the
LUNA collaboration, the $\Iso{14}{N}(p,\gamma)\Iso{15}{O}$ bottleneck
reaction of the CNO-cycle \citep{lunan14:04}. The influence on
globular cluster age determinations has already been discussed by
\citet{icfbb:04}, and we will briefly comment on this. Our main
interest, however, in 
this bottleneck reaction is again how stars of low and intermediate
mass might be influenced in their evolution.

The outline of this paper is as follows:
After describing briefly our stellar
evolution code and the specific reaction rates for the
$\tralpha$ process and the $\Iso{14}{N}$ reaction in Sect.~2, we 
present the results for the three stellar evolution cases described above
in Sect.~3, followed by the conclusion in the final section.

\newpage

\section{Stellar evolution program}

\subsection{Basic properties}

All calculations were done with the Garching
stellar evolution code as described by \citet{wsch:2000}. The program
was modified in only minor aspects since then. To summarize briefly 
the main physical ingredients, the code incorporates the OPAL equation of
state \citep{rsi:96} and the OPAL opacity tables \citep{ir:96}
supplemented by the molecular opacities of
\citet{af:94}. In the calculations presented here, the equation of
state (EOS) of Irwin \citep[see][]{csirwin:2003} has been used,
which is based on
the OPAL EOS. Microscopic diffusion of all elements is implemented and has been
included in the calculations of Sect.~\ref{s:pop3}. Convection is
treated according to standard mixing length theory with the
Schwarzschild criterion for stability.
Mass loss is included according to Reimers' formula \citep{reimers},
in the generalized form \citep{ir:83} containing
a free scaling parameter $\eta$, which we will specify for
each case.

Nuclear burning is taken into account by a network which treats
hydrogen and helium burning separately, unless both protons and
$\alpha$-particles are present and temperatures are high enough for
helium processing. In such cases (see Sect.~\ref{s:pop3}) the whole
network and any mixing process are treated simultaneously. For details
on this see \citet{scsw:2001}.

\subsection{$\nbottle$  and $3\,\alpha$ rate}

The standard implementation of the first rate\footnote{The term ``rate'' is
 to be understood as $N_A^{(n-1)} \langle \sigma v \rangle$, that is
 as the Maxwellian-averaged reaction rate \citep[see][]{nacre:99} in units of
 $\mathrm{cm^3} \, \mathrm{mol^{-1}} \, \mathrm{s^{-1}}$ for $n$ reaction partners.}
 is according to the
recommendation in \citet[Table~VI; hereafter referred to as
 Adel98]{adel:98}, with the astrophysical
$S(0)$-factor being $3.5$~keV~b. NACRE \citep{nacre:99} gives 
$S(0)=3.2\pm 0.8$~keV~b. The new LUNA value \citep{lunan14:04} is
$1.7\pm 0.1\,\mathrm{(stat)} \pm 0.2\,\mathrm{(sys)}$~keV~b; $S'(0)$
 and $S''(0)$ have been left unchanged at the 
Adelberger values and the new rate therefore differs from the old one
 by a constant factor. This is in
agreement with the procedure of \citet{icfbb:04}.


The standard implementation of the $3\,\alpha$ rate in our program is
the analytic form of \citet[CFHZ85]{CFHZ:85}. Since
\citet{tralpha:05} used the analytic fit to the NACRE  
rate for reference, we implemented the same rate for
the comparisons. 
The new ISOL rate was
available to us in tabulated form (Chr.~Diget, private
communication). 
Fig.~\ref{f:1} shows all $\tralpha$ rates compared with the NACRE
analytic rate. 
Note that in the temperature range of interest
to us the new rate lies within the uncertainty of the NACRE rate
\citep{nacre:99} and that the analytic fit of the
NACRE rate deviates from the tabulated values by up to 5\%.
The ISOL rate agrees quite well
with that of \citet{CFHZ:85} over an extended range below temperatures of
several billion degree. 
The critical temperature above which
helium burning is taken into account was
set to $5\cdot 10^7$~K; using still lower values has no
influence on the models.

\section{Sample calculations}

\subsection{Core helium flash in a Pop.~II star}
\label{s:hefl}

In this section we present the results of our sample calculations and 
compare them  with reference calculations that  use the NACRE  and Adel98
rates for the $\tralpha$ and $\nbottle$ reaction rates,
respectively. We begin with the onset of helium burning in stars
of low mass and degenerate helium cores, the so-called core helium
flash. 
We performed calculations for two masses,  $0.8\,\msun$
and $1.0\,\msun$, and three metallicities, $Z=0.001,\, 0.0001,\,
0.00001$ and initial helium content $Y=0.25$. No mass loss was
assumed. The calculations were started from the
zero-age main sequence (ZAMS) and followed through the flash
until the star settles on the horizontal branch (HB) and then up to
the early AGB. 

\begin{figure}
\centerline{\includegraphics[draft=false,scale=0.53]{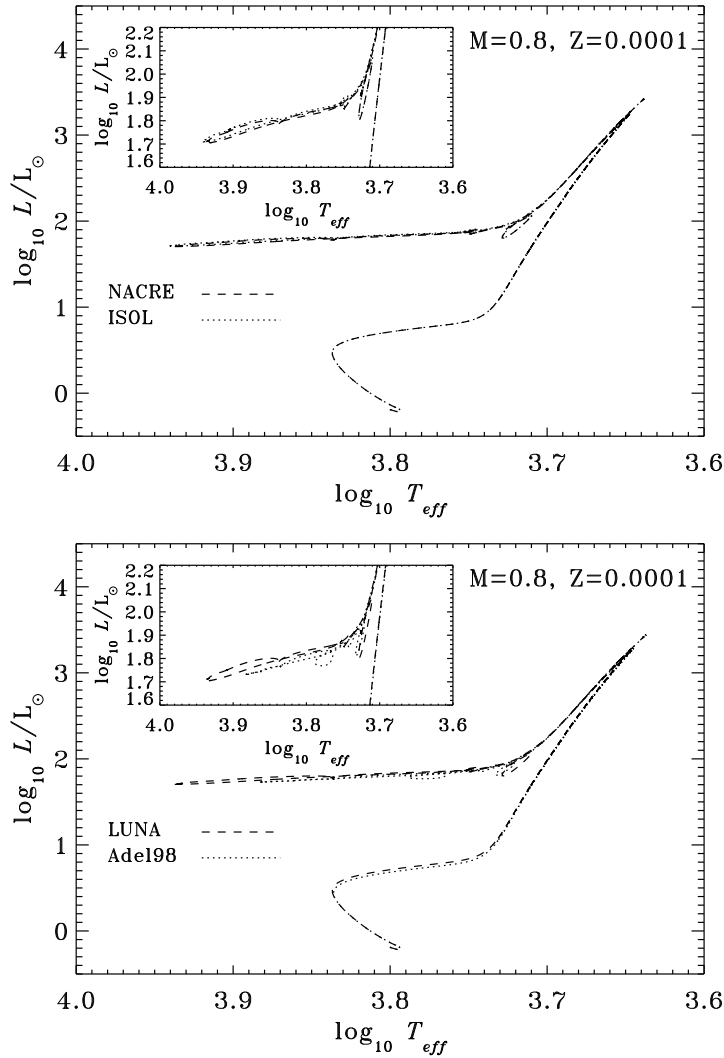}}
\caption[]{Evolution of a stellar model of $0.8\, \msun$
and $Z=10^{-4}$ from ZAMS to HB. In the upper panel we varied the
$\tralpha$- and in the lower one the $\nbottle$-rate. The insets show
the approach to the HB in greater detail.} 
\protect\label{f:2}
\end{figure}

Updating the $\tralpha$ rate, the changes are very small but as 
would be expected
from a slightly lower rate. The tip of the RGB is almost unchanged, the
luminosity being 2\% higher for the new rate.
The core mass increases very little, between $0.002$ and 
$0.003\, \msun$ depending on mass and metallicity. The age of the
models also increases by a mere 0.2~Myr at most. We show the resulting
tracks in the Hertzsprung-Russell Diagram (HRD) in Fig.~\ref{f:2} for
the lowest metallicity and mass, and list a few quantities in Table~\ref{t:1}. Note
that even the secondary flashes during the settling to the HB are almost
indistinguishable. The agreement between the two cases is as close for
all other cases. 

We then varied the $\nbottle$ rate ($\tralpha$ = NACRE). The effect turned
out to be much larger, 
and can be seen in Fig.~\ref{f:2} (lower panel). We notice differences already
around the turn-off, and in particular during the approach to the
HB; the effective temperature of the zero-age HB model increases by up
to 20\% for the LUNA rate. 

\begin{table}
\caption{Selected properties of one of our Pop.~II models
($M=0.8\,\msun$; $Z=10^{-4}$) at critical stages of
its evolution from ZAMS to ZAHB evolved with
various combinations of reaction rates as shown in
Fig.~\ref{f:2}. $M_\mathrm{He}$ is the helium core mass in solar
unit at the helium flash.} 
\protect\label{t:1}
\begin{center}	
\begin{tabular}{ccccc}
\hline
$^{14}$N+p: & & Adel98 & \multicolumn{2}{c}{LUNA} \\
\hline
3$\alpha$: & & NACRE & NACRE & ISOL \\
\hline 
Turn off & 
$\begin{array}{c} {\log L/L_\odot}   \\ {\log T_{\rm eff} } \\
  {\rm Age [Myr]} \end{array}$ &

$\begin{array}{c} 0.437 \\ 3.8360 \\ 11130 \end{array}$ &

$\begin{array}{c} 0.467 \\ 3.8368 \\ 11273 \end{array}$ &

$\begin{array}{c} 0.467 \\ 3.8368 \\ 11273 \end{array}$
\\
\hline 
RGB-tip & 
$\begin{array}{c} {\log L/L_\odot }  \\ {\log T_{\rm eff} } \\
  {M_\mathrm{He}} \\ {\rm Age [Myr]} \end{array}$ &

$\begin{array}{c} 3.285 \\ 3.6449 \\ 0.4990 \\ 12638 \end{array}$ &

$\begin{array}{c} 3.237 \\ 3.6479 \\ 0.5052 \\ 12699 \end{array}$ &

$\begin{array}{c} 3.246 \\ 3.6474 \\ 0.5074 \\ 12699 \end{array}$
\\
\hline 
ZAHB & 
$\begin{array}{c} {\log L/L_\odot }  \\ {\log T_{\rm eff} } \\
\end{array}$ & 

$\begin{array}{c} 1.744 \\ 3.8614 \\ \end{array}$ &

$\begin{array}{c} 1.715 \\ 3.9324 \\ \end{array}$ &

$\begin{array}{c} 1.722 \\ 3.9344 \\ \end{array}$
\\
\hline
\end{tabular}
\end{center}
\end{table}

\begin{figure}[h]
\centerline{\includegraphics[draft=false,angle=90,scale=0.50]{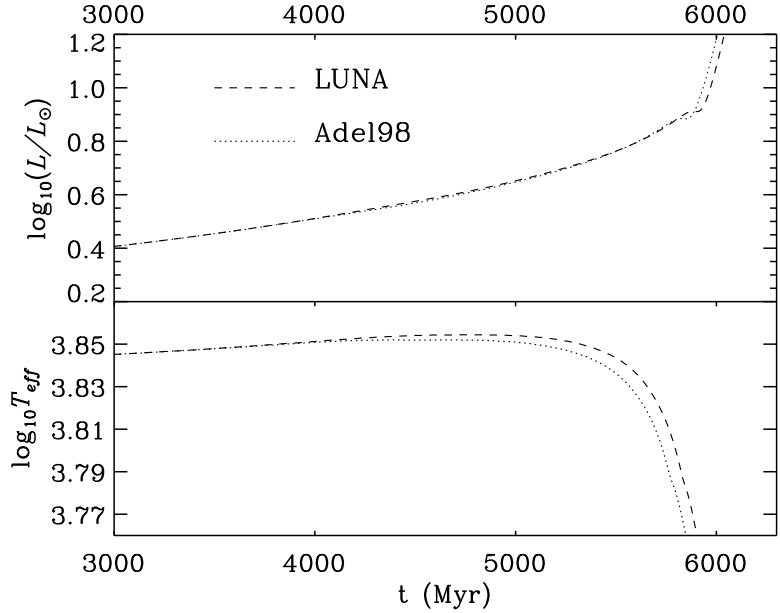}}
\caption[]{Comparison of luminosity and effective temperature as
functions of time (in Myr) for a star with $1\,\msun$ and $Z=0.001$,
calculated with both the Adelberger (dotted line) and LUNA (dashed line)
$\nbottle$ rate.} 
\protect\label{f:4}
\end{figure}

At this point it is worthwhile to clarify the result found by
\citet{icfbb:04}. They showed that due to the improved $\nbottle$ rate
the ages of globular clusters increase by up to 1~Gyr. 
This is not simply due to the slower bottleneck reaction leading to 
longer timescale for hydrogen fusion. 
In Fig.~\ref{f:4} we display for one example ($1\,\msun$, $Z=0.001$) the
evolution of luminosity and effective temperature during 
main-sequence and red-giant-branch evolution. Evidently, the luminosity
evolution on the main sequence is hardly affected by the nuclear
rate. Energy generation takes place within a core
that has a temperature only marginally higher but is more extended
as the result of a shallower temperature profile. So, while at the
center hydrogen is burning at a slower pace (same $T$, lower reaction
rate), more mass is involved in the nuclear fusion. The time it takes
to deplete hydrogen in the core is therefore somewhat higher (in our
case +190~Myr), but the exhausted core is larger ($+0.013\,M_\odot$). 
Due to that larger core mass, 
towards the end of the main sequence the central 
temperature is higher by approximately 5\% for the LUNA rate. Another
consequence of the different core evolution is a modified  
temperature gradient, and thus $\teff$ at the end of the main
sequence evolves differently (Fig.~\ref{f:4}), influencing the turn-off
luminosity which is defined as the value of $L$ at the bluest point
along the MS track. To be more quantitative, the turn-off in the
case of the Adelberger rate is at $t=4.38$~Gyr, $\log\teff = 3.852$, and $\log
L/\lsun = 0.554$. For the revised LUNA rate the hottest point is
reached later, at $\log\teff = 3.854$, when $\log L/\lsun = 0.593$ and
$t=4.62$~Gyr, an increase of 240~Myr. Note, however, that at the
turn-off age of the Adelberger case, $\teff = 3.854$ and $\log L/\lsun =
0.558$ for LUNA, i.e., almost identical. 
Thus, the change in the turn-off is due to the modified
interior structure, which influences the morphology of the evolutionary track. 
We have also investigated a few other cases and find that at even lower
metallicity ($Z=0.0001$) the age differences decrease to 1-2\%. 

\newpage

\subsection{Core helium flash in a Pop.~III
star}\label{s:pop3}

A particularly interesting variant of the core helium flash is that 
found in
metal-free (Pop.~III) stars. Details on this subject 
were provided by
\citet{wcschs:2000} and 
\citet{scsw:2001}. 
We have calculated the evolution of a star of $M=1\,\msun$, $Y_i =
0.23$ ($Z=0$). 
We varied both rates as before, the reference case using 
the \citet{adel:98} $\nbottle$ and NACRE $\tralpha$ rates. Some
properties of the models are listed in
Table~\ref{t:2}. We also investigated the case ``LUNA + CFHZ85'', but
do not discuss it further, because it is indistinguishable from the ``LUNA +
ISOL'' case.

The first
particularity of Pop.~III evolution is a loop in the HRD after the main
sequence, which is due to spurious carbon production in the hot core
and the ignition of the CNO-cycle \citep[for details
see][]{wcschs:2000}. The presence of this loop depends on the stellar
mass; it is present in our reference case, but vanishes in all cases 
with the new (lower) LUNA rate for $\nbottle$.
As the reason for the disappearance of the loop we identify
the fact that the
lower CNO-luminosity prevents the creation of a convective core, which
otherwise leads to an increase of the hydrogen abundance in the core and
thus to a luminosity enhancement. 
We note that the creation of the
first in-situ carbon nuclei is not influenced by the choice of the
$\tralpha$ rate.

\begin{table}[h]
\caption{Selected properties of Pop.~III models evolved with
various combinations of reaction rates. We
list luminosity and helium core mass at the onset
of the first and second core helium flash, as well as the maximum
helium luminosity ($\log L_{\rm He}^{\rm max}/L_\odot
$) and the mass at which the maximum temperature is reached ($M_{\rm
T_{\rm max}}/M_\odot$) at the time of $\log L_{\rm He}^{\rm
max}/L_\odot$. In addition, the surface abundances (mass fractions)
after the  flash-induced mixing event is given for key elements.}
\protect\label{t:2}
\begin{center}
\begin{tabular}[6]{llccc} \hline \hline
$^{14}$N+p && Adel98 & \multicolumn{2}{c}{LUNA} \\ \hline
$3\alpha$ && NACRE & NACRE & ISOL \\ \hline \hline
MS & CN-loop & yes & no & no \\ \hline
& $\log(L/L_\odot)$ & 2.324 & 2.302 & 2.311 \\
& $\log(L_{\rm He}^{\rm max}/L_\odot)$ & 10.08 & 10.01 &
10.03 \\
\rbox{1$^{\rm st}$ flash} & $M_{\rm He}/M_\odot$ & 0.4784 & 
0.4755 & 0.4780 \\
& $M_{\rm T_{\rm max}}/M_\odot$ & 0.1849 & 0.1818 &
0.1863 \\ \hline
& H & 0.5288 & 0.5369 & 0.5366 \\
& He & 0.4582 & 0.4510 & 0.4510 \\
dredge up & $^{12}$C & 0.00417 & 0.00292 & 0.00302 \\
surface & $^{13}$C & 0.00111 & 0.00080 & 0.00083 \\
abdcs. &  $^{14}$N & 0.00772 & 0.00837 & 0.00861 \\
& $^{15}$N & $2.6\cdot 10^{-7}$ & 
$1.1\cdot 10^{-7}$ & $1.1\cdot 10^{-7}$ \\
& $^{16}$O & $3.9\cdot 10^{-6}$ &
$3.4\cdot 10^{-6}$ & $4.0\cdot 10^{-6}$ \\ \hline
& $\log(L/L_\odot)$ & 3.532 & 3.487 & 3.498 \\
& $\log(L_{\rm He}^{\rm max}/L_\odot)$ & 9.592 & 9.764 & 9.770 \\
\rbox{2$^{\rm nd}$ flash} & $M_{\rm He}/M_\odot$ & 0.4585 & 0.4615
& 0.4638 \\
& $M_{\rm T_{\rm max}}/M_\odot$ & 0.0789 & 0.0665 & 0.0729 \\
\hline\hline 
\end{tabular}
\end{center}
\end{table}

The second particular event is the (first) core helium flash that
happens at a much lower luminosity in Pop.~III stars than in those
with $Z \gtrsim 10^{-6}$, due to the higher core temperature. As the H-shell 
temperatures are higher as well, the so-called entropy barrier is
lower and a mixing event between the hydrogen-rich envelope and the
hot carbon-enriched helium layers is possible \citep[see
also][]{FII:00}. As a consequence, the 
envelope will be enriched drastically in CNO-products and helium (see
Table~\ref{t:2}). 
As in the case of ordinary Pop.~II
stars, the new $\tralpha$ rate has only a very moderate influence on
the flash: it starts at slightly higher luminosity and core mass, but
the increase is not significant. The peak helium luminosity $L_
\mathrm{He}^\mathrm{max}$ is unchanged, within the computational uncertainties.

In contrast, the new
$\nbottle$ rate has a much stronger influence. The 
flash luminosity is lower due to the higher shell
temperatures and the core mass is higher. This is probably a
consequence of the fact that the core mass-luminosity relation
depends on the temperature exponent of the shell hydrogen burning. For
CNO-burning, it is $L \propto M_c^7$, and for $pp$-burning $\propto M_c^3$. Very
metal-poor stars have a very small contribution from the CNO-cycle
only and therefore follow closely the latter relation. A lower
$\nbottle$ rate leads to a smaller temperature exponent, and thus
the core mass is slightly higher at given luminosity.

The result of the flash-induced dredge-up shows larger (relative)
variations (Table~\ref{t:2}), but with 
no apparent systematics, except that in the
table the dredge-up appears generally to be stronger for the older
rates, while for the latest rate more CN-burning could take place. 

After the star has settled again on the giant branch, it
resumes its evolution since core helium burning has been
extinguished as a result of the first flash and mixing. A second, more
moderate flash appears at a standard RGB tip luminosity ($\log L/\lsun
\approx 3.5$). Table~\ref{t:2} demonstrates again that the largest
changes are due to the new LUNA $\nbottle$ rate, leading to lower
ignition luminosity but higher helium core mass. 
The largest variation found in Table~\ref{t:2} is therefore
between the reference case (col.~3) and that updating the $\nbottle$ rate only
(col.~4). 

\subsection{Helium burning and thermal pulses in a Pop.~I star}
\label{s:agb}

As a third case we have investigated that of a typical
solar-metallicity intermediate-mass star evolving into and through the
Asymptotic Giant Branch (AGB) phase. The composition of our model was
$X=0.695$, $Y=0.285$ and $Z=0.02$. Mass loss according to Reimers was taken
into account with the scaling parameter $\eta$ being 0.4 on the RGB and 0.5 on
the AGB.

\begin{figure}
\centerline{\includegraphics[draft=false,scale=0.50,angle=90]{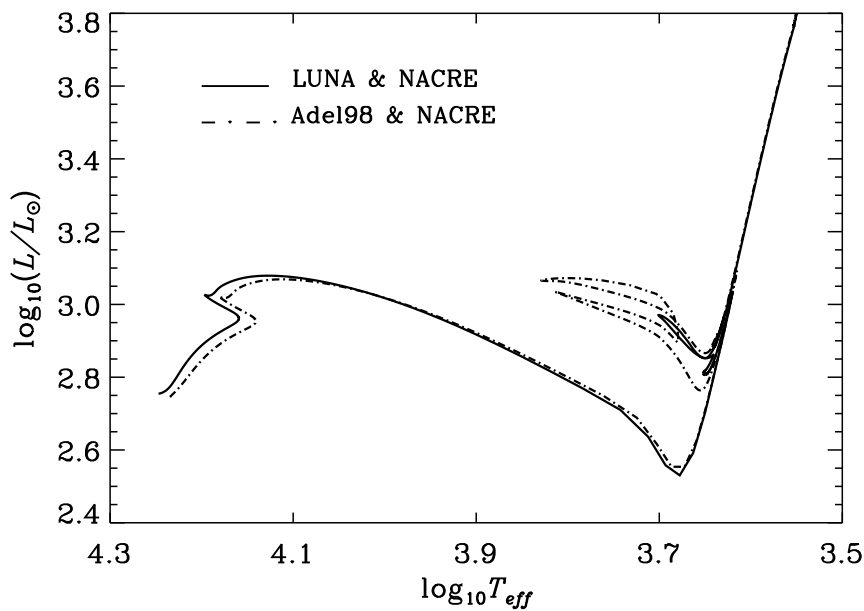}}
\caption[]{Evolution in the HRD for the Pop.~I model with
$M=5\,\msun$, showing the influence of changing the $\nbottle$
reaction rate. The two cases are indicated in the figure.} 
\protect\label{f:6}
\end{figure}

In Fig.~\ref{f:6} we show the evolution of this model in the HRD for
the two $\nbottle$ reaction rates. Using the updated LUNA
rate the main-sequence evolution takes place at higher
temperatures, implying a more compact structure of the model. This is
similar to the Pop.~II model 
(Sect.~3.1) but more pronounced. As the
evolution of intermediate mass stars is very sensitive to the internal
composition profile, the consequences for the core helium burning
phase are quite large: the blue loop gets significantly
shorter. 

\begin{figure}
\centerline{\includegraphics[draft=false,scale=0.50,angle=90]{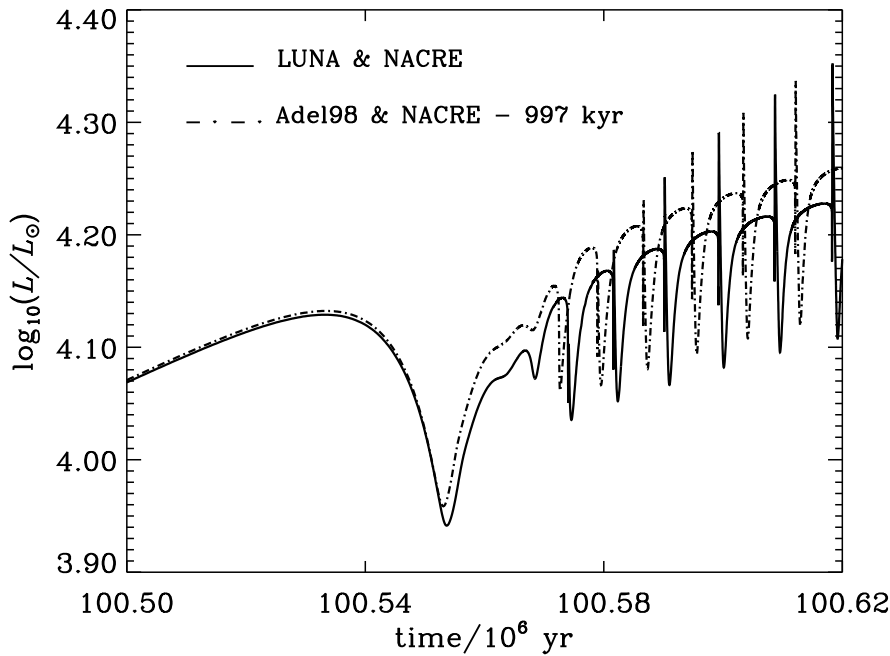}}
\caption[]{Thermal pulses on the AGB for two different $\nbottle$
reaction rates. Shown is the total luminosity 
as function of time (in million years). To display better the
differences the time axis for the Adel98 $\nbottle$  case has been
shifted by -997~kyr.}
\protect\label{f:7}
\end{figure}

The luminosity variations during the thermal pulses (TP) along the AGB
are shown in Fig.~\ref{f:7}. The TPs start earlier by about 1~million
years for the LUNA rate, which is due to a shorter main-sequence
lifetime. We have therefore shifted the time axis for the Adel98 case to
coincide at the pre-flash luminosity maximum. The pulse behaviour is
similar, although the peak luminosity and peak helium luminosity (not
shown) are
higher, and the interpulse duration longer for the new LUNA rate. This
is consistent with earlier findings concerning the influence of
H-burning rates on pulse behaviour \citep{ds:76}, and in particular with the recent
investigation by \citet{ha:2004}, who found stronger flashes for lower
$\nbottle$ rates in a $2\,\msun$ model.  These differences become
smaller with increasing 
pulse number.  We followed 15 (LUNA) respectively 8 (Adel98) pulses
and then stopped the calculations.

Using the new LUNA $\nbottle$ reaction rate, we then compared
models calculated with the three variants of the $\tralpha$ rate. 
Again, the blue 
loops, being most sensitive to structure variations, show the clearest
reaction, and get even shorter for the new ISOL rate.
Concerning the thermal pulses, they are extremely similar to each
other, and if shifted by  -80~kyr for the CFHZ85 and -67~kyr for the new ISOL
$\tralpha$ rate relative to the NACRE case they completely agree.

Finally, we mention the chemical composition of all cases investigated
in terms of central carbon and oxygen abundance after core helium
burning. The carbon abundance ranges from 0.200 to
0.223 with the highest value reached for the older reaction
rates. Consequently the core most rich in oxygen is obtained for the LUNA
$\nbottle$ and new $\tralpha$ rates combined. However, in spite of
noticeable variations, they are minor compared with the uncertainties
still present due to the $\Iso{12}{C}(\alpha,\gamma)\Iso{16}{O}$
rate. As an example, we quote here the result of \citet[their Tab.~6]{bccmpt:2000}
that the time spent in the blue loop by a $5\,\msun$ star changes by up
to 2.12 Myr (33\%), the mass of the CO-core by 
$0.26\,\msun$ (5\%), and the C/O ratio by a factor of 3 when varying
the rate within a factor of 2.3, which is comparable to the rate
uncertainty \citep{kunz:2002}.

\section{Conclusions}

We have investigated the influence of new  rates for the $\nbottle$
and $\tralpha$ reactions on the evolution of low- and intermediate-mass
stars, considering cases in which helium burning proceeds under
thermally unstable conditions.
In all cases we find negligible changes in the evolution and also
in the interior evolution due to the new $\tralpha$ rate. The RGB tip
brightness is slightly increased due to the lower rate at very low
temperatures corresponding to the earliest phases of helium ignition.
The largest effect shows up
in the blue loops during core helium burning of the
$5\,\msun$ star, emphasizing the sensitivity of these loops to
details of the interior structure.

The $\nbottle$ rate has a definitely stronger
influence. It prolongs the duration of central
hydrogen burning, increases the turn-off temperature and
thus indirectly the turn-off location, it leads to a disappearance of
the CN flash in the Pop.~III post-main-sequence star, strongly
reduces the blue loop during core helium burning of a $5\,\msun$ star,
and also influences the thermal  
pulses. 
Interestingly, these are, apart from a very minor shift in
time, almost identical in the case of varying the $\tralpha$ rate.

We conclude that the new $\tralpha$ rate has no influence on the
evolution of low and intermediate-mass stars, and that the effect of the 
LUNA $\nbottle$ is tiny, but noticeable. Further investigations into
its effect on more massive stars as well as of the effect 
of the $\tralpha$ rate on
massive stars are indicated. 
If the differences in the various rate
determinations we have used are representative of the
experimental errors, then both rates are no longer a source of
uncertainty for stellar modeling.

\begin{acknowledgements}
We thank Chr.~Diget for making available the rate table.
F.~Meissner kindly provided data for the turn-off values of metal-poor
stars, and F.~Herwig information about his AGB-models with the new
rates. A.W.\ is grateful to L.~Girardi for explanations concerning the core
mass-luminosity relation.
A.M.S\ is supported in part by NSF grant PHY-0070928.
\end{acknowledgements}
\bibliographystyle{aa}
\bibliography{tralph}

\end{document}